\documentstyle[12pt]{article}
\begin{document}
\title{{\Large New approach to }$\varepsilon ${\Large -entropy and Its compa
rison
with Kolmogorov's }$\varepsilon ${\Large -entropy}}
\author{Kei Inoue$^{*}$, Takashi Matsuoka$^{*}$ and
Masanori Ohya\thanks{Email: ohya@is.noda.sut.ac.jp}\\
Department of Information Sciences\\
Science University of Tokyo\\
Noda City, Chiba 278, Japan}
\date{}
\maketitle

running head: New approach to $\varepsilon $-entropy

\bigskip

the name and mailing adress of the authors:
\vspace{2mm}\\
\hspace*{10mm} Prof. Masanori Ohya \\
\hspace*{10mm} Department of Information Sciences\\
\hspace*{10mm}  Science University of Tokyo\\
\hspace*{10mm}  Noda City, Chiba 278, Japan
\hspace*{10mm}  ohya@is.noda.sut.ac.jp
\newpage
\begin{abstract}
Kolmogorov introduced a concept of $\varepsilon $-entropy to analyze
information in classical continuous system. The fractal dimension of
geometrical sets was introduced by Mandelbrot as a new criterion to analyze
the complexity of these sets. The $\varepsilon $-entropy and the fractal
dimension of a state in general quantum system were introduced by one of the
present authors in order to characterize chaotic properties of general
states.

In this paper, we show that $\varepsilon$-entropy of a state includes
Kolmogorov $\varepsilon$-entropy, and the fractal dimension of a state
describe fractal structure of Gaussian measures.
\end{abstract}

\section{Introduction}

The $\varepsilon $-entropy was introduced by Kolmogorov (1963) using
the mutual entropy with respect to two random variables $f$ and $g$. The
entropy $S(f)$ of a random variable $f$ is usually infinite on a continuous
probability space. On the other hand, the $\varepsilon $-entropy $S_{{\rm %
Kolmogorov}}(f;\varepsilon )$ ( $S_{{\rm K}}(f;\varepsilon )$ for short )
can be bounded. Therefore, we can use the $\varepsilon $ -entropy to analyze
random variables in classical system. This $\varepsilon $-entropy $S_{{\rm K}%
}(f;\varepsilon )$ expresses a degree of information transmission in the $%
\varepsilon $-neighborhood of a random variable $f$.

By the way, Mandelbrot introduced a new criterion to analyze complexity of
geometrical sets, it is so called fractal dimension (Mandelbrot, 1982), which is
different from the euclidean dimensions. Usual fractal theory mostly treats
only geometrical sets. It is desirable to extend the fractal dimensions in
order to characterize some other objects. One of the present authors
introduced the notion of $\varepsilon $- entropy $S_{{\rm Ohya}}(\mu
;\varepsilon )$ ($S_{{\rm O}}(\mu ;\varepsilon )$ for short) for a state in
order to formulate the fractal dimension of a state in general quantum
system (GQS for short) (Ohya, 1989),(Ohya, 1991),(Ohya and Petz, 1993).
Actually, the capacity dimensions ,
which is one of the fractal dimensions for geometrical sets, was given by
the $\varepsilon $-entropy. Namely, it is defined by
\begin{equation}
d_{C}(X)=\lim_{\varepsilon \to 0}\frac{\log N_{X}(\varepsilon )}{\log \frac{1%
}{\varepsilon }},
\end{equation}
where $N_{X}(\varepsilon )$ is the minimum numbers of a convex set with
diameter $\varepsilon $ covering a set $X$ and $\log N_{X}(\varepsilon )$ is
called $\varepsilon $-entropy of a geometrical set $X$
(Kolmogorov and Tihomirov, 1961). The capacity dimension characterize fractal
structure of the geometrical sets as a limiting behavior of $\varepsilon $-e
ntropy
when $\varepsilon $ approach to $0$.

Our fractal dimension of a state is formulated by extending the concept of
the capacity dimension to GQS. That is, the fractal dimension of a state is
expressed by the $\varepsilon $-entropy of a state instead of the $%
\varepsilon $-entropy of a geometrical set $X$ and characterize fractal
structure of a state.

These $\varepsilon $-entropy and fractal dimension provide new criteria
describing the complexity of states, so that they can be used to distinguish
two states even when they have the same value for the entropy. For instance,
we could analyze the complexity for some systems by these criteria
(Ohya, 1991),(Matsuoka and Ohya,1995),(Akashi, 1992).

In this paper, we examine the similarity and difference of two $\varepsilon$%
-entropies $S_{{\rm K}}$ and $S_{{\rm O}}$ for Gaussian measures (states) on
a Hilbert space. It is shown that our $\varepsilon$ -entropy and fractal
dimension are useful to classify Gaussian measures, in the case of that
Kolmogorov's $\varepsilon$-entropy can not be used.

\section{Kolmogorov's $\varepsilon$-entropy for random variables}

In this section, we remind the definition of Kolmogorov's $\varepsilon $
-entropy for random variables. Let $(\Omega ,\Im ,\mu )$ be a probability
space and $M(\Omega )$ be the set of all random variables, and $f,g$ be two
random variables on $\Omega $ with valued on a metric space $(X,d)$. Let $%
\mu _{f}$ be a probability measure associated with a random variable $f$.
Then the mutual entropy $I(f,g)$ of the random variable $f$ and $g$ is
defined by (Gelfand and Yaglom, 1959)
\begin{eqnarray*}
I(f,g) &=&S\left( \mu _{fg},\mu _{f}\otimes \mu _{g}\right) \\
&=&\left\{
\begin{tabular}{ll}
${\int_{X\times X}}\frac{d\mu _{fg}}{d\mu _{f}\otimes \mu _{g}}\log \frac{%
d\mu _{fg}}{d\mu _{f}\otimes \mu _{g}}d\mu _{f}\otimes \mu _{g}$ & $%
\hspace*{\fill}(\mu _{fg}\ll \mu _{f}\otimes \mu _{g})$ \\
$\infty $ & $\hspace*{\fill}($otherwise$)$%
\end{tabular}
,\right.
\end{eqnarray*}
\noindent where $S(\cdot ,\cdot )$ is relative entropy (Kulback-Leibler
information), $\mu _{f}\otimes \mu _{g}$ is the direct product probability
measure of $f$ and $g$, and $\mu _{fg}$ is the joint probability
distribution of $f$ and $g$, $\frac{d\mu _{fg}}{d\mu _{f}\otimes \mu _{g}}$
is the Radon-Nikodym derivative of $\mu _{fg}$ with respect to $\mu
_{f}\otimes \mu _{g}$. Moreover the entropy $S(f)$ of the random variable $f$
is given by
\begin{equation}
S(f)=I(f,f).
\end{equation}
$S(f)$ is often infinite in continuous case. Kolmogorov introduced the $%
\varepsilon $- entropy for a random variable $f$ as follows;
\begin{equation}
S_{{\rm K}}(f;\varepsilon )=\inf_{g}\left\{ I(f,g);g\in M_{d}(f;\varepsilon
)\right\} ,
\end{equation}
where
\begin{equation}
M_{d}(f;\varepsilon )=\left\{ g\in M(\Omega )\ ;\ \sqrt{\int_{X\times
X}d(x,y)^{2}d\mu _{fg}(x,y)}\leq \varepsilon \right\} .
\end{equation}

\section{$\varepsilon$-entropy and fractal dimensions of a state}

The $\varepsilon $-entropy and the fractal dimension of a state were
introduced in Ohya (1991) for GQS. In this section, we review these
formulations in the framework of classical measure theory.

In the information theory, an input state is described by a state
(probability measure in continuous classical system, density operator in
usual quantum system) and it is sent to an output system (receiver) through
some channel denoted by $\Lambda^{*}$. A channel is a transmitter (e.g.
optical fiber), mathematically it is a mapping from an input state space to
an output state space.

When an input state $\mu$ dynamically changes to an output state $\bar{\mu}%
(\equiv \Lambda^{*}\mu)$ under a channel $\Lambda^{*}$, we ask how much
information carried by $\mu$ can be transmitted to the output state through
the channel $\Lambda^{*}$. It is the mutual entropy that represents this
amount of information transmitted from $\mu$ to $\bar{\mu}$. Hence, the
mutual entropy depends on an input state and a channel. In this scheme, the
mutual entropy is formulated as follows:

Let $(\Omega_{1},\Im_{1})$ be an input space, $(\Omega_{2},\Im_{2})$ be an
output space and $P(\Omega_{k})$ be the set of all probability measures on $%
(\Omega_{k},\Im_{k})(k=1,2)$. We can call the following linear mapping $%
\Lambda^{*}$ from $P(\Omega_{1})$ to $P(\Omega_{2})$ a channel (Markov
kernel):
\begin{eqnarray}
\bar{\mu}(Q) & = & \Lambda^{*}\mu(Q)  \nonumber \\
{} & = & \int_{\Omega_{1}} \lambda(\omega,Q) \; d\mu(\omega) \hspace{5mm}\mu
\in P(\Omega_{1}),  \nonumber
\end{eqnarray}
where $\lambda$ is a mapping from $\Omega_{1} $$\times$$\Im_{2}$ to $[0,1]$
satisfying the following conditions: \vspace{3mm}

\begin{itemize}
\item[(1)]  $\lambda (\cdot ,Q)$ is a measurable function on $\Omega _{1}$
for each $Q$ $\in \Im _{2}$.

\item[(2)]  $\lambda (\omega ,\dots )\in P(\Omega _{2})$ for each $\omega
\in \Omega _{1}$.
\end{itemize}

\vspace*{3mm} \noindent The compound state $\Phi$ of $\mu$ and $\bar{\mu}$
is given by
\begin{equation}
\Phi(Q_{1} \times Q_{2})=\int_{Q_{1}} \lambda(\omega,Q_{2}) \; d\mu(\omega),
\end{equation}
for any $Q_{1} $$\in $$\Im_{1},Q_{2}$$\in $$\Im_{2}$. The mutual entropy in
classical continuous system can be expressed by the relative entropy $%
S(\cdot , \cdot)$ of the compound state $\Phi$ and the direct product state $%
\Phi_{0}=\mu \otimes \Lambda^{*}\mu$: \vspace{3mm}

\noindent $I(\mu ;\Lambda ^{*})=S(\Phi ,\Phi _{0})$ ($\equiv I(\mu ,\bar{\mu}%
;\Phi )$)
\begin{equation}
=\left\{
\begin{array}{cl}
{\int_{\Omega _{1}\times \Omega _{2}}}\frac{d\Phi }{d\Phi _{0}}\log \frac{%
d\Phi }{d\Phi _{0}}\;d\Phi _{0} & (\Phi \ll \Phi _{0}) \\
\infty & ({\rm otherwise})
\end{array}
\right. ,
\end{equation}
where $\frac{d\Phi }{d\Phi _{0}}$ is the Radon-Nikodym derivative of $\Phi $
with respect to $\Phi _{0}$.

In the following discussion, we consider the case of $(\Omega_{1},$$\Im_{1})$
$=(\Omega_{2},$$\Im_{2}) $$\equiv $ $(\Omega,$$\Im)$ for simplicity.

We shall give the definition of Kolmogorov's $\varepsilon $-entropy $S_{{\rm %
K}}(\mu ;\varepsilon )$ for a general probability measure $\mu $ on $(\Omega
,\Im )$.
\begin{equation}
S_{{\rm K}}(\mu ;\varepsilon )\equiv \inf \left\{ I(\mu ,\bar{\mu};\Phi
)\;;\;||\mu -\bar{\mu}||\le \varepsilon \right\} ,
\end{equation}
where $\Vert \mu \Vert $ is a certain norm of $\mu $.

The $\varepsilon $-entropy of a state $\mu $ is defined as follows (Ohya, 1989)%
. \vspace{3mm}

{\em Definition1: (The }$\varepsilon ${\em -entropy of a state }$\mu \in
P\left( \Omega \right) ${\em )}
\begin{equation}
S_{{\rm O}}(\mu ;\varepsilon )\equiv \inf_{\Lambda ^{*}}\left\{ J(\mu
;\Lambda ^{*});\Vert \mu -\Lambda ^{*}\mu \Vert \leq \varepsilon \right\} ,
\end{equation}
where $\Vert \mu \Vert $ is a certain norm of $\mu $ and
\begin{equation}
J(\mu ;\Lambda ^{*})\equiv \sup_{\Gamma ^{*}}\left\{ I(\mu ;\Gamma
^{*});\Gamma ^{*}\mu =\Lambda ^{*}\mu \right\} ,
\end{equation}
Here $J(\mu ;\Lambda ^{*})$ is called the maximum mutual entropy w.r.t. $\mu
$ and $\Lambda ^{*}$.

The $\varepsilon$-entropy $S_{{\rm O}}(\mu; \varepsilon)$ is a bit more
general than the Kolmogorov $\varepsilon$-entropy $S_{{\rm K}}(f;
\varepsilon)$ for random variables, more precisely the $\varepsilon$-entropy
$S_{{\rm O}}(\mu;\varepsilon)$ is different from $S_{{\rm K}}(\mu;\varepsilon)$
in the following points.\vspace{2mm}

\begin{itemize}
\item[(1)]  The definition is based on states (probability measures) not
only random variables.

\item[(2)]  Several possibilities to choose the norm of states.

\item[(3)]  The concept of the maximum mutual entropy $J(\mu ;\Lambda ^{*})$
is used, which is a very essential as is described in section 6.
\end{itemize}

\ \vspace{-2mm}

This fractal dimension of a state $\mu $ in a classical continuous system is
defined by the $\varepsilon $-entropy of states. \vspace{2mm}

{\em Definition2: (The capacity dimension of a state }$\mu \in P\left(
\Omega \right) ${\em \ )}
\begin{equation}
d_{{\rm C}}^{{\rm O}}\equiv \lim_{\varepsilon \to 0}\frac{S_{{\rm _{O}}}(\mu
;\varepsilon )}{\log \frac{1}{\varepsilon }}
\end{equation}

\section{Gaussian measure and Gaussian channel on a Hilbert Space}

We briefly review the Gaussian communication processes treated by Baker et
al (Baker, 1978),(Yanagi, 1988).

Let ${\cal B}$ be the Borel $\sigma $-field of a real separable Hilbert
space ${\cal H}$ and $\mu $ be a Borel probability measure on ${\cal B}$
satisfying
\begin{equation}
\int_{{\cal H}}\Vert x\Vert ^{2}d\mu (x)<\infty .
\end{equation}
Further, we denote the set of all positive self-adjoint trace class
operators on ${\cal H}$ by $T({\cal H})_{+}$ $\left( \equiv \{R\in {\cal B}(%
{\cal H});\;R\geq 0,\right. $ $\left. R=R^{*},{\rm tr}R<\infty \}\right) $
and define the mean vector $m_{\mu }$$\in $ ${\cal H}$ and the covariance
operator $R_{\mu }$$\in $$T({\cal H})_{+}$ of $\mu $ such as
\begin{equation}
\langle x_{1},m_{\mu }\rangle =\int_{{\cal H}}\langle x_{1},y\rangle d\mu
(y),
\end{equation}
\begin{equation}
\langle x_{1},R_{\mu }x_{2}\rangle =\int_{{\cal H}}\langle x_{1},y-m_{\mu
}\rangle \langle y-m_{\mu },x_{2}\rangle d\mu (y),
\end{equation}
for any $x_{1},x_{2},y\in {\cal H}$. A Gaussian measure $\mu $ in ${\cal H}$
is a Borel measure such that for each $x$$\in $${\cal H}$, there exist real
numbers $m_{x}$ and $\sigma _{x}(>0)$ satisfying
\begin{eqnarray*}
\lefteqn{\mu \{y\in {\cal H};\langle y,x\rangle \leq a\}} \\
&=&\int_{-\infty }^{a}{\frac{1}{\sqrt{2\pi }\sigma _{x}}}\exp \left\{ \frac{%
-(t-m_{x})^{2}}{2\sigma _{x}^{2}}\right\} dt.
\end{eqnarray*}
The notation $\mu =[m,R]$ means that $\mu $ is a Gaussian measure on ${\cal H%
}$ with a mean vector $m$ and a covariance operator $R$.

Let $({\cal H}_{1},{\cal B}_{1})$ be an input space, $({\cal H}_{2},{\cal B}
_{2})$ be an output space and $P_{G}^{(k)}$ be the set of all Gaussian
probability measures on $({\cal H}_{k},{\cal B}_{k})(k=1,2)$. We consider
the case of $({\cal H}_{1},{\cal B}_{1})=({\cal H}_{2},{\cal B}_{2}) \equiv
( {\cal H},{\cal B})$ for simplicity. Moreover, let $\mu \in P({\cal H})$ be
a Gaussian measure of the input space and $\mu_{0} \in P({\cal H})$ be a
Gaussian measure indicating a noise of the channel. Then, a Gaussian channel
$\Lambda^{*}$ from $P({\cal H})$ to $P({\cal H})$ is defined by the
following mapping $\lambda :{\cal H} \times {\cal B} \rightarrow [0,1]$ such
as
\begin{equation}
\bar{\mu}(Q)=\Lambda^{*}\mu(Q) \equiv \int_{{\cal H}} \lambda(x,Q)d\mu(x)
\end{equation}
\begin{equation}
\lambda(x,Q) \equiv \mu_{0}(Q^{x}),
\end{equation}
\begin{equation}
Q^{x} \equiv \left\{ y \in {\cal H};Ax+y \in Q \right\},x \in {\cal H},Q \in
{\cal B},
\end{equation}
where {\it A} is a linear transformation from ${\cal H}$ to ${\cal H}$ and $%
\lambda$ satisfies the following conditions: \vspace{3mm}

\begin{itemize}
\item[(1)]  $\lambda (x,\cdot )\in P({\cal H})$ for each fixed $x\in {\cal H}
$,

\item[(2)]  $\lambda (\cdot ,Q)$ is measurable function on $({\cal H},{\cal B%
})$ for each fixed $Q\in {\cal B}$.
\end{itemize}

\vspace*{3mm} \noindent The compound measure $\Phi$ derived form the input
measure $\mu$ and the output measure $\bar{\mu}$ is given by
\begin{equation}
\Phi(Q_{1} \times Q_{2}) = \int_{Q_{1}} \lambda(x,Q_{2})d\mu(x)
\end{equation}
for any $Q_{1},Q_{2} \in {\cal B}$.

In particular, let $\mu $ be $[0,R]\in P({\cal H})$ and $\mu _{0}$ be $%
[0,R_{0}]\in P({\cal H})$. Then, output measure $\Lambda ^{*}\mu =\bar{\mu}$
can be expressed as
\begin{equation}
\Lambda ^{*}\mu =[0,ARA^{*}+R_{0}].
\end{equation}
When the dimension of ${\cal H}$ is finite, the mutual entropy (information)
with respect to $\mu $ and $\Lambda ^{*}$ become \cite{GY}
\begin{equation}
I(\mu ;\Lambda ^{*})=\frac{1}{2}\log \frac{|ARA^{*}+R_{0}|}{|R_{0}|}\,,
\end{equation}
where $|$ $ARA^{*}$ $+R_{0}$ $|$, $|R_{0}$ $|$ are determinants of $ARA^{*}$
$+R_{0}$, $R_{0}$.

\section{$\varepsilon$-entropy and fractal dimension of a state for Gaussian
measures in the random variable norm}

The $\varepsilon $-entropy of states described in section 3 is different
from Kolmogorov's definition of the $\varepsilon $-entropy for random
variable. In this section, we show that two definitions coincide when ${\cal %
H}={\bf R}^{n}$ and the norm of a state $\mu _{f}$ is defined by
\begin{equation}
\Vert \mu _{f}\Vert =\sqrt{\frac{1}{n}\sum_{i=1}^{n}\int_{\Omega
}|f_{i}|^{2}d\mu }.
\end{equation}
Then the distance between two states $\mu _{f}$ and $\mu _{g}$ induced by
the above norm leads
\begin{equation}
\Vert \mu _{f}-\mu _{g}\Vert =\sqrt{\frac{1}{n}\sum_{i=1}^{n}\int_{\Omega
}|f_{i}-g_{i}|^{2}d\mu }.
\end{equation}
We call this norm random variable norm (R. V. norm for short) in the sequel.
In this paper, we only consider Gaussian measures with the mean $0$ and
Gaussian channels.

Let an input state $\mu_{f}=[0,R]$ be induced from a $n$-dimensional random
vector $f=(f_{1},$ $\ldots,$ $f_{n})$ and its output state $%
\Lambda^{*}\mu_{f}$ be denoted by $\mu_{g}$, where $g$ is random vector $g=$
$(g_{1},$ $\ldots,$ $g_{n})$ induced from $\Lambda^{*}$. \vspace{3mm}

{\em Lemma1:} If the distance of two states is given by the above R.V. norm,
then
\begin{equation}
J(\mu _{f}\ ;\ \Lambda ^{*})=I(\mu _{f}\ ;\ \Lambda ^{*})
\end{equation}

{\em proof:} From the assumption, the Gaussian channel $\Lambda ^{*}$ is
represented by a conditional probability density of $g$ with respect to $f$
as
\begin{eqnarray}
p(y|x) &=&\frac{1}{(2\pi )^{\frac{n}{2}}\sqrt{|R_{0}|}}  \nonumber \\
{} &&\hspace{3mm}\times \exp \left\{ -\frac{1}{2}(y-Ax)R_{0}^{-1}(y-Ax)^{t}%
\right\}  \nonumber \\
{} &&\hspace{2.5cm}x,y\in {\bf R}^{n},  \nonumber
\end{eqnarray}
where $R_{0}$ is the covariance matrix associated to the channel $\Lambda
^{*}$.Then the compound state $\Phi $ of $\mu _{f}$ and $\Lambda ^{*}\mu
_{f}=\mu _{g}$ is equal to the joint probability measure $\mu _{fg}=[0,C]$
of $f$ and $g$ such as

\begin{eqnarray*}
&&\mu _{fg}(Q_{1}\times Q_{2}) \\
&=&\int_{Q_{1}\times Q_{2}}\frac{1}{(2\pi )^{n}\sqrt{|C|}}\exp \left\{ -%
\frac{1}{2}zC^{-1}z^{t}\right\} dz\quad Q_{1},Q_{2}\in {\cal B}({\bf R}^{n}),
\end{eqnarray*}
where $z$ is the $2n$-dimensional random vector $(x,y)$ $=(x_{1},$$\ldots ,$$%
x_{n},$ $y_{1},$ $\ldots ,$ $y_{n})$ and $C$ is the following covariance
matrix of $\mu _{fg}$ ;
\begin{equation}
C=\left(
\begin{array}{cc}
R & RA^{t} \\
AR & ARA^{t}+R_{0}
\end{array}
\right) ,  \label{5.1.1}
\end{equation}
where $R,$ $RA^{t},$ $AR,$ $ARA^{t}$$+R_{0}$ are $n$$\times n$ matrix and $%
(R)_{ij}=E(f_{i}f_{j}),$ $(RA^{t})_{ij}$ \\$=E(f_{i}g_{j}), \;(AR)_{ij}$
$=E(g_{i}f_{j}),\;(ARA^{t}$$+R_{0})_{ij}=E(g_{i}g_{j})$ for each $(i,j) $
$(i,j$\\$=1,\ldots ,n)$.

For the channel $\Gamma^{*}$ satisfying $\Lambda^{*}\mu_{f}$$=\Gamma^{*}
\mu_{f},$ $\Gamma^{*}\mu_{f}$ is a $n$-dimensional Gaussian measure $\mu_{h}$
induced from a $n$-dim\-ensional random vector $h=(h_{1},$ $\ldots,$$h_{n})$%
, so that we have
\begin{eqnarray*}
\lefteqn{J(\mu_{f};\Lambda^{*}) } \\
& = & \sup_{\Gamma^{*}}\left\{I(\mu_{f};\Gamma^{*}); \Lambda^{*}\mu_{f}
=\Gamma^{*}\mu_{f}\right\} \\
{} & = & \sup_{\Gamma^{*}}\left\{I(\mu_{f};\Gamma^{*}); \Vert
\Lambda^{*}\mu_{f}-\Gamma^{*}\mu_{f} \Vert =0 \right\} \\
{} & = & \sup_{h}\left\{I(\mu_{f};\Gamma^{*}); \Vert \mu_{g}-\mu_{h} \Vert
=0\right\} \\
{} & = & \sup_{h}\left\{I(f,h);E[d(g,h)^{2}] =0\right\} \\
{} & = & \sup_{h}\left\{I(f,h); E[d(g_{i},h_{i})^{2}] =0 \;(i=1,\ldots,n)
\right\} \\
{} & = & \sup_{h}\left\{I(f,h); g_{i}=h_{i} \; a.e. \;\;
(i=1,\ldots,n)\right\}\,.
\end{eqnarray*}
From $g_{i}=h_{i} \; a.e. \;\;(i=1,\ldots,n)$, we obtain
\begin{equation}
\mu_{fg}=\mu_{fh}\,.
\end{equation}
Therefore,
\begin{equation}
I(\mu_{f};\Lambda^{*})=I(f,h)=I(\mu_{f};\Gamma^{*}),
\end{equation}
which implies
\begin{equation}
J(\mu_{f};\Lambda^{*})=I(\mu_{f};\Lambda^{*}) \,.
\end{equation}

Using the above lemma, the following theorem holds. \vspace{3mm}

{\em Theorem1: }Under the same assumption as Lemma 1

\begin{itemize}
\item[(1)]  $S_{{\rm O}}(\mu _{f};\varepsilon )=S_{{\rm K}}(f;\varepsilon )$
$={\frac{1}{2}\sum_{i=1}^{n}\log \max \left( \frac{\lambda _{i}}{\theta ^{2}}%
,1\right) }\,,$
\end{itemize}

where $\lambda _{1},\ldots ,$$\lambda _{n}$ are the eigenvalues of $R$ and
$\theta ^{2}$ is a constant uniquely determined by the equation ${%
\sum_{i=1}^{n}}$$\min (\lambda _{i},$$\theta ^{2})$ $=\varepsilon ^{2}$.

\begin{itemize}
\item[(2)]  $d_{{\rm C}}^{{\rm O}}(\mu _{f})=n$.
\end{itemize}

{\em proof: } (1) Let $\bar{{\cal C}}$ be the set of all Gaussian channels
from ${{\cal B}({\bf R}^{n})}$ to ${{\cal B}({\bf R}^{n})}$ and $\bar{{\cal C%
}}(\mu _{f};\varepsilon )$ be the set of all Gaussian channels from ${\cal B}%
({\bf R}^{n})$ to ${\cal B}({\bf R}^{n})$ satisfying $\Vert \mu _{f}-\Lambda
^{*}\mu _{f}\Vert \leq \varepsilon $. According to Lemma 1, we obtain
\begin{eqnarray*}
S_{{\rm O}}(\mu _{f}\;;\varepsilon ) &=&\inf \left\{ J(\mu _{f};\Lambda
^{*});\Lambda ^{*}\in \bar{{\cal C}}(\mu _{f};\varepsilon )\right\} \\
{} &=&\inf \left\{ I(\mu _{f};\Lambda ^{*});\Lambda ^{*}\in \bar{{\cal C}}%
(\mu _{f};\varepsilon )\right\} \,,
\end{eqnarray*}
From( \ref{5.1.1}), we have
\begin{eqnarray*}
S_{{\rm O}}(\mu _{f}\;;\varepsilon ) &=&\inf \left\{ I(\mu _{f}\;;\Lambda
^{*});\Lambda ^{*}\in \bar{{\cal C}}(\mu _{f};\varepsilon )\right\} \\
&=&\inf \left\{ I(f,g);\mu _{fg}\in \bar{{\cal S}}(\mu _{f};\varepsilon
)\right\} \\
&=&S_{{\rm K}}(f;\varepsilon ),
\end{eqnarray*}
where $\bar{{\cal S}}(\mu _{f};\varepsilon )=\left\{ \mu _{fg};\sqrt{\int_{%
{\bf R}^{n}\times {\bf R}^{n}}d(x,y)^{2}d\mu _{fg}(x,y)}\leq \varepsilon
\right\} $. \newline
The expression of the $\varepsilon $-entropy ${\frac{1}{2}\sum_{n=1}^{n}\log
\left\{ \max \left( \frac{\lambda _{i}}{\theta ^{2}},1\right) \right\} }$
was obtained by \\ Pinsker (1963). \vspace{5mm} \newline
(2) Since $S_{{\rm O}}(\mu _{f}\;;\varepsilon )={\frac{1}{2}\sum_{n=1}^{n}}%
\log {\left\{ \max \left( \frac{\lambda _{i}}{\theta ^{2}},1\right) \right\}
}$, we have
\begin{eqnarray*}
d_{{\rm C}}^{{\rm O}}(\mu _{f}) &=&d_{{\rm C}}^{{\rm K}}(\mu
_{f})=\lim_{\varepsilon \to 0}\frac{S_{{\rm O}}(\mu _{f};\varepsilon )}{\log
{\frac{1}{\varepsilon }}} \\
{} &=&\lim_{\varepsilon \to 0}\frac{\frac{1}{2}\sum_{i=1}^{n}\log \left\{
\max \left( \frac{\lambda _{i}}{\theta ^{2}},1\right) \right\} }{\log \frac{1%
}{\varepsilon }} \\
&&{\rm \hspace{3cm}}\left( i.e.\sum_{i=1}^{n}\min (\lambda _{i},\theta
^{2})=\varepsilon ^{2}\right) \\
{} &=&\lim_{\varepsilon \to 0}\frac{\frac{1}{2}\sum_{i=1}^{n}\log \frac{%
\lambda _{i}}{\theta ^{2}}}{\log \frac{1}{\varepsilon }}\;\;\;\left(
\sum_{i=1}^{n}\theta ^{2}=\varepsilon ^{2}\right) \\
{} &=&\lim_{\varepsilon \to 0}\frac{\frac{1}{2}\sum_{i=1}^{n}\log \frac{%
n\lambda _{i}}{\varepsilon ^{2}}}{\log \frac{1}{\varepsilon }}=n\,.
\end{eqnarray*}
In this case, our $\varepsilon $-entropy coincides with Kolmogorov's $%
\varepsilon $-entropy and the fractal dimension of the state $\mu _{f}$ is
identical to the dimension of Hilbert space.

\section{$\varepsilon$-entropy and fractal dimension of a state for Gaussian
measures in the total variation norm}

In the following discussion, we only consider the case of $\dim ({\cal H})=1$%
, that is, ${\cal H}={\bf R}$, the state $\mu =[0,\sigma ^{2}]$ is a
one-dimensional Gaussian measure (distribution) and the distance of two
states is given by the total variation norm.

In this section, we show that our $\varepsilon $-entropy traces to the
fractal property of a Gaussian measure but Kolmogorov's does not. The
difference between $S_{{\rm O}}$ and $S_{{\rm K}}$ come from the norm of
measures taken. We take the norm by the total variation, namely,
\begin{equation}
\Vert \mu \Vert =|\mu |({\bf R})
\end{equation}
Let ${\cal H}_{1}={\cal H}_{2}={\bf R}$. Then $A$ becomes a real number $%
\beta $ and the noise of a channel is exhibited by one-dimensional Gaussian
measure $\mu _{0}=[0,\sigma _{0}^{2}]\in P({\cal H})$, so that the output
state $\Lambda ^{*}\mu $ is represented by $[0,\beta ^{2}\sigma ^{2}+\sigma
_{0}^{2}]$. We calculate the maximum mutual entropy in the following two
cases ; (1) $\beta ^{2}\sigma ^{2}+\sigma _{0}^{2}\geq \sigma ^{2}$, (2) $%
\beta ^{2}\sigma ^{2}+\sigma _{0}^{2}<\sigma ^{2}$ . Since the channel $%
\Lambda ^{*}$ depends on $\beta $ and $\sigma _{0}^{2}$, we put $\Lambda
^{*}=\Lambda _{(\beta ,\sigma _{0}^{2})}^{*}$ . As the density function for
Gaussian measures are error functions, we first give an order estimation for
the difference of two Gaussian measures.

\smallskip \medskip {\em Lemma2:} If $\beta ^{2}\sigma ^{2}+\sigma
_{0}^{2}\geq \sigma ^{2}$ and $\Vert \mu -\Lambda _{(\beta ,\sigma
_{0}^{2})}^{*}\mu \Vert $$=$$|\mu -$ $\Lambda _{(\beta ,\sigma
_{0}^{2})}^{*}\mu |({\bf R})=\delta $, then
\vspace{1mm}\newline
(1) ${\frac{4}{\sqrt{2\pi }}}{\frac{\sqrt{\beta ^{2}\sigma ^{2}+\sigma
_{0}^{2}}-\sigma }{\sigma }}=\delta +o(\delta )$, \newline
(2) $\left\{ \Lambda _{(\beta ,\sigma _{0}^{2})}^{*};\Vert \mu -\Lambda
_{(\beta ,\sigma _{0}^{2})}^{*}\mu \Vert \leq \varepsilon \right\} =$

$\left\{ \Lambda _{(\beta ,\sigma _{0}^{2})}^{*};\frac{4}{\sqrt{2\pi }}{%
\frac{\sqrt{\beta ^{2}\sigma ^{2}+\sigma _{0}^{2}}-\sigma }{\sigma }}=\delta
+o(\delta ),\;\delta \in M(\varepsilon )\right\} $, \vspace{3mm}\newline
where $o(\delta )$ is an order of $\delta $ : $\lim_{\delta \to 0}$ $%
o(\delta )$$=0$ and $M(\varepsilon )=$
$\left\{ \delta \in {\bf R};0\leq \delta \leq \varepsilon \right\} $.

{\em  proof: } (1) Let $p_{1},p_{2}$ be the density functions
of $\mu ,\Lambda ^{*}\mu $, respectively. Then, we have
\begin{equation}
\Vert \mu -\Lambda _{(\beta ,\sigma _{0}^{2})}^{*}\mu \Vert =\int_{{\bf R}%
}|p_{1}(x)-p_{2}(x)|dx  \label{6.1.1}
\end{equation}
Since $p_{i}\;(i=1,2)$ are even functions, we obtain
\begin{eqnarray}
\lefteqn{\int_{{\bf R}}|p_{1}(x)-p_{2}(x)|dx}   \\
{} &=&2\int_{0}^{\infty }|p_{1}(x)-p_{2}(x)|dx   \\
{} &=&2\left( \int_{0}^{a}(p_{1}(x)-p_{2}(x))dx+\int_{a}^{\infty
}(p_{2}(x)-p_{1}(x))dx\right)   \\
{} &=&4\int_{0}^{a}(p_{1}(x)-p_{2}(x))dx  \label{6.1.2} \\
&&\;\;\left( a=\sqrt{\left( \frac{1}{\sigma ^{2}}-\frac{1}{\beta ^{2}\sigma
^{2}+\sigma _{0}^{2}}\right) ^{-1}\log \frac{\beta ^{2}\sigma ^{2}+\sigma
_{0}^{2}}{\sigma ^{2}}}\right)  \nonumber \\
{} &\leq &\frac{4a}{\sqrt{2\pi }}\left( \frac{1}{\sigma }-\frac{1}{\sqrt{%
\beta ^{2}\sigma ^{2}+\sigma _{0}^{2}}}\right)   \\
{} &\leq &\frac{4}{\sqrt{2\pi }}\frac{\sqrt{\beta ^{2}\sigma ^{2}+\sigma
_{0}^{2}}-\sigma }{\sigma },
\end{eqnarray}
where $a$ is a real number satisfying $p_{1}(a)=p_{2}(a)$, and the first
inequality is led by a geometrical approximation of the equation (\ref{6.1.2}%
) and the second inequality is obtained from the inequality $\log x\leq x-1$
for any positive number $x$.

Since $\Vert \mu - \Lambda^{*}_{(\beta,\sigma_{0}^{2})}\mu \Vert$ and $\frac{%
4}{\sqrt{2\pi}} \frac{\sqrt{\beta^{2}\sigma^{2}+\sigma_{0}^{2}} -\sigma}{%
\sigma}$ are monotone decreasing for \\ $\beta^2\sigma^2+\sigma_{0}^2 \to
\sigma^{2}$ and
\begin{eqnarray*}
\lefteqn{\lim_{\beta^{2}\sigma^{2}+\sigma_{0}^{2} \to \sigma^2} ||\mu -
\Lambda^*_{(\beta, \sigma_0^2)}\mu|| } \\
{} & = & \lim_{\beta^2\sigma^2+\sigma_0^2 \to \sigma^2} \frac{4}{\sqrt{2\pi}}%
\frac{\sqrt{\beta^{2}\sigma^{2}+\sigma_{0}^{2}} -\sigma}{\sigma} \\
{} & = & 0\;,
\end{eqnarray*}
the above inequality implies
\begin{equation}
\Vert \mu-\Lambda^{*}_{(\beta,\sigma_{0}^{2})}\mu \Vert = \delta
\Leftrightarrow \frac{4}{\sqrt{2\pi}} \frac{\sqrt{\beta^{2}\sigma^{2}+
\sigma_{0}^{2}}-\sigma}{\sigma}= \delta+o(\delta) ,
\end{equation}
where $o(\delta)$ is an order of $\delta$.

\noindent (2) From Lemma2(1), \vspace{3mm}\newline
$\left\{ \Lambda _{(\beta ,\sigma _{0}^{2})}^{*};\Vert \mu -\Lambda _{(\beta
,\sigma _{0}^{2})}^{*}\mu \Vert =\delta \right\} $
\begin{equation}
=\left\{ \Lambda _{(\beta ,\sigma _{0}^{2})}^{*};\frac{4}{\sqrt{2\pi }}\frac{%
\sqrt{\beta ^{2}\sigma ^{2}+\sigma _{0}^{2}}-\sigma }{\sigma }=\delta
+o(\delta )\right\} .  \label{6.1.3}
\end{equation}
Let $M(\varepsilon )$ be the set of all $\delta \in {\bf R}$ satisfying $%
0\leq \delta \leq \varepsilon $. From (\ref{6.1.3}), it is clear that \vspace{%
3mm}\newline
$\left\{ \Lambda _{(\beta ,\sigma _{0}^{2})}^{*};\Vert \mu -\Lambda _{(\beta
,\sigma _{0}^{2})}^{*}\mu \Vert \leq \varepsilon \right\} =$
\begin{equation}
\left\{ \Lambda _{(\beta ,\sigma _{0}^{2})}^{*};\frac{4}{\sqrt{2\pi }}\frac{%
\sqrt{\beta ^{2}\sigma ^{2}+\sigma _{0}^{2}}-\sigma }{\sigma }=\delta
+o(\delta ),\;\;\;\delta \in M(\varepsilon )\right\}
\end{equation}

{\em Lemma3: }Let $\Lambda _{\delta (\beta ,\sigma _{0}^{2})}^{*}$ be a
channel satisfying $\beta ^{2}\sigma ^{2}+\sigma _{0}^{2}\geq \sigma ^{2}$
and $\Vert \mu -\Lambda _{\delta (\beta ,\sigma _{0}^{2})}^{*}\mu \Vert
=\delta $ for any $\delta \in M(\varepsilon )$. If a Gaussian channel $%
\Lambda _{\delta (\beta ,\sigma _{0}^{2})}^{*}$ satisfies the condition $%
\beta ^{2}\leq \frac{C_{\delta }-\delta }{\sigma ^{2}}$, then we have
\begin{equation}
J(\mu ;\Lambda _{\delta (\beta ,\sigma _{0}^{2})}^{*})=\frac{1}{2}\log \frac{%
1}{\delta }+\frac{1}{2}\log \sigma ^{2}\left( 1+\frac{\sqrt{2\pi }}{4}%
(\delta +o(\delta ))\right) ^{2},
\end{equation}
where $C_{\delta }=\beta ^{2}\sigma ^{2}+\sigma _{0}^{2}$ is a constant
determined by $\Vert \mu -\Lambda _{\delta (\beta ,\sigma _{0}^{2})}^{*}\mu
\Vert $$=\delta $.

{\em proof:} The mutual entropy of $\mu $ with respect to channel $\Lambda
^{*}$ is
\begin{equation}
I(\mu ;\Lambda ^{*})=\frac{1}{2}\log \frac{\beta ^{2}\sigma ^{2}+\sigma
_{0}^{2}}{\sigma _{0}^{2}}.
\end{equation}
\noindent Thus, if $\Lambda _{\delta (\beta ,\sigma _{0}^{2})}^{*}$ is any
channel satisfying $\Vert \mu -\Lambda _{\delta (\beta ,\sigma
_{0}^{2})}^{*}\mu \Vert =\delta $, then we have from the above lemma 2
\vspace{3mm}
\begin{eqnarray}
\lefteqn{I(\mu ;\Lambda _{\delta (\beta ,\sigma _{0}^{2})}^{*})}  \nonumber
\\
&=&\frac{1}{2}\log \frac{1}{C_{\delta }-\beta ^{2}\sigma ^{2}}+\frac{1}{2}%
\log \sigma ^{2}\left( 1+\frac{\sqrt{2\pi }}{4}(\delta +o(\delta ))\right)
^{2},  \nonumber
\end{eqnarray}
The assumption implies,
\begin{eqnarray*}
\lefteqn{J(\mu ;\Lambda ^{*})} \\
&=&\sup_{\Lambda _{(\bar{\beta},\bar{\sigma _{0}^{2}})}^{*}}\left\{ I(\mu
;\Lambda _{(\bar{\beta},\bar{\sigma _{0}^{2}})}^{*});\Lambda _{\delta (\beta
,\sigma _{0}^{2})}^{*}\mu =\Lambda _{\delta (\bar{\beta},\bar{\sigma _{0}^{2}%
})}^{*}\mu \right\} \\
&=&\sup_{\Lambda _{(\bar{\beta},\bar{\sigma _{0}^{2}})}^{*}}\left\{ I(\mu
;\Lambda _{(\bar{\beta},\bar{\sigma _{0}^{2}})}^{*});\Vert \Lambda _{\delta
(\beta ,\sigma _{0}^{2})}^{*}\mu -\Lambda _{\delta (\bar{\beta},\bar{\sigma
_{0}^{2}})}^{*}\mu \Vert =0\right\} \\
&=&\sup_{\Lambda _{(\bar{\beta},\bar{\sigma _{0}^{2}})}^{*}}\left\{ I(\mu
;\Lambda _{(\bar{\beta},\bar{\sigma _{0}^{2}})}^{*});\beta ^{2}\sigma
^{2}+\sigma _{0}^{2}=\bar{\beta}^{2}\sigma ^{2}+\bar{\sigma}_{0}^{2}\right\}
\\
&=&\sup_{\bar{\beta}}\left\{ \frac{1}{2}\log \frac{1}{C_{\delta }-\bar{\beta}%
^{2}\sigma ^{2}}+\frac{1}{2}\log \sigma ^{2}\cdot f(\delta )\;\;;\bar{\beta}%
^{2}\leq \frac{C_{\delta }-\delta }{\sigma ^{2}}\right\} \\
&=&\frac{1}{2}\log \frac{1}{\delta }+\frac{1}{2}\log \sigma ^{2}\cdot
f(\delta )
\end{eqnarray*}
\vspace*{3mm} where $\bar{\beta}^{2}=\frac{C_{\delta }-\delta }{\sigma ^{2}}$
and $f(\delta )=\left( 1+\frac{\sqrt{2\pi }}{4}(\delta +o(\delta ))\right)
^{2}$.

{\em Lemma4: }Let $\Lambda _{\delta (\beta ,\sigma _{0}^{2})}^{*}$ be a
channel satisfying $\beta ^{2}\sigma ^{2}+\sigma _{0}^{2}<\sigma ^{2}$ and $%
\Vert \mu -\Lambda _{\delta (\beta ,\sigma _{0}^{2})}^{*}\mu \Vert =\delta $
for any $\delta \in M(\varepsilon )$. If a Gaussian channel $\Lambda
_{\delta (\beta ,\sigma _{0}^{2})}^{*}$ satisfies the condition $\beta
^{2}\leq \frac{C_{\delta }-\delta }{\sigma ^{2}}$, then we have
\begin{equation}
J(\mu ;\Lambda _{(\beta ,\sigma _{0}^{2})}^{*})=\frac{1}{2}\log \frac{1}{%
\delta }+\frac{1}{2}\log \frac{\sigma ^{2}}{\left( 1+\frac{\sqrt{2\pi }}{4}%
(\delta +o(\delta ))\right) ^{2}}
\end{equation}
where $C_{\delta }=\beta ^{2}\sigma ^{2}+\sigma _{0}^{2}$ is a constant
determined by $\Vert \mu -\Lambda _{\delta (\beta ,\sigma _{0}^{2})}^{*}\mu
\Vert $$=\delta $.

\ \vspace{2mm}\newline
{\em proof:} Similarly proved as Lemma 2 and Lemma 3. \vspace{3mm}\newline
Using the above those lemmas, we obtain the following theorem. \vspace{3mm}

{\em Theorem2: }Under the same conditions of Lemma 3 and 4, we have

\begin{itemize}
\item[(1)]  $S_{{\rm O}}(\mu ;\varepsilon )=\frac{1}{2}\log \frac{1}{%
\varepsilon }+\frac{1}{2}\log \frac{\sigma ^{2}}{\left( 1+\frac{\sqrt{2\pi }%
}{4}(\varepsilon +o(\varepsilon ))\right) ^{2}}>S_{{\rm K}}(\mu ;\varepsilon
)=0$

\item[(2)]  $d_{{\rm C}}^{{\rm O}}(\mu )=\frac{1}{2}$
\end{itemize}

\ \newline
{\em proof:} (1) From Lemma 3 and 4, we have
\begin{eqnarray*}
\lefteqn{S_{{\rm O}}(\mu ;\varepsilon )=\inf_{\Lambda ^{*}}\left\{ J(\mu
;\Lambda ^{*});\Vert \mu -\Lambda ^{*}\mu \Vert \leq \varepsilon \right\} }
\\
{} &=&\inf_{\delta }\left\{
\begin{array}{l}
{\frac{1}{2}\log \frac{1}{\delta }+\frac{1}{2}\log \sigma ^{2}\left( 1+\frac{%
\sqrt{2\pi }}{4}(\delta +o(\delta ))\right) ^{2}} \\
{\frac{1}{2}\log \frac{1}{\delta }+\frac{1}{2}\log \frac{\sigma ^{2}}{\left(
1+\frac{\sqrt{2\pi }}{4}(\delta +o(\delta ))\right) ^{2}}}
\end{array}
\hbox{\quad ;\quad }\delta \in M(\varepsilon )\right\} \\
{} &=&\inf_{\delta }\left\{ \frac{1}{2}\log \frac{1}{\delta }+\frac{1}{2}%
\log \frac{\sigma ^{2}}{\left( 1+\frac{\sqrt{2\pi }}{4}(\delta +o(\delta
))\right) ^{2}}\hbox{\quad ;\quad }\delta \in M(\varepsilon )\right\} \\
{} &=&\frac{1}{2}\log \frac{1}{\varepsilon }+\frac{1}{2}\log \frac{\sigma
^{2}}{\left( 1+\frac{\sqrt{2\pi }}{4}(\varepsilon +o(\varepsilon ))\right)
^{2}},
\end{eqnarray*}
\noindent because $\frac{1}{2}\log \frac{1}{\delta }+\frac{1}{2}\log \frac{%
\sigma ^{2}}{\left( 1+\frac{\sqrt{2\pi }}{4}(\delta +o(\delta ))\right) ^{2}}
$ is monotone decreasing with respect to $\delta $ \vspace{5mm} . \newline
(2) From (1), we obtain
\begin{eqnarray*}
d_{{\rm C}}^{{\rm O}}(\mu ) &=&{\lim_{\varepsilon \to 0}\frac{S_{{\rm O}%
}(\mu ;\varepsilon )}{\log \frac{1}{\varepsilon }}} \\
&=&\lim_{\varepsilon \to 0}\frac{{\frac{1}{2}\log \frac{1}{\varepsilon }+%
\frac{1}{2}\log \frac{\sigma ^{2}}{\left( 1+\frac{\sqrt{2\pi }}{4}%
(\varepsilon +o(\varepsilon ))\right) ^{2}}}}{{\log \frac{1}{\varepsilon }}}
\\
&=&\frac{1}{2}\;\;.
\end{eqnarray*}
This result show that (1) the fractal dimension of a Gaussian measure $\mu $
in the total variation norm describe fractal structure of Gaussian measures,
and (2) The fractal dimension of a Gaussian measure $\mu $ is always $0$ if
we use the Kolmogorov $\varepsilon $-entropy.

We concluded that our fractal dimension of states is a new criterion to
study a chaotic aspect of Gaussian measures.

\end{document}